\documentclass[twocolumn]{aastex631}
\usepackage{float}
\usepackage{graphicx}
\usepackage{amsmath}
\usepackage{natbib}
\usepackage{color}
\usepackage{verbatim}
\usepackage[utf8]{inputenc}
\usepackage{mathtools}
\usepackage{amssymb}
\usepackage{textcomp}
\usepackage{enumitem}
\usepackage{braket}
\usepackage{graphicx} % Required for inserting images

\begin{document}
\title{A Unified View of Stellar and AGN Feedback across Galactic Mass Scales}

\author[0009-0004-2298-284X]{Allan VanZandt}
\affiliation{Independent researcher, Washington, DC, 20009, USA}
\affiliation{Department of Astronomy, University of 
Michigan, 323 West Hall, 1085 S. University Ave., Ann Arbor, MI, 48105-1107, USA} 

\author[0000-0002-5564-9873]{Eric F. Bell}
\affiliation{Department of Astronomy, University of 
Michigan, 323 West Hall, 1085 S. University Ave., Ann Arbor, MI, 48105-1107, USA} 

\begin{abstract}

The baryon lifting hypothesis posits that the feedback energy required in a galaxy reflects the gravitational binding energy of the baryons in its circumgalactic medium. Both stellar feedback from supernovae and active galactic nucleus (AGN) feedback from supermassive black hole (SMBH) accretion drive this regulation, yet are typically studied separately via scaling relations linking stellar mass ($M_{\star}$), black hole mass ($M_{\mathrm{BH}}$), and halo mass ($M_{\mathrm{halo}}$), often limited by consistent $M_{\mathrm{halo}}$ estimates across wide mass ranges. Using the globular cluster system mass--halo mass relation to estimate $M_{\mathrm{halo}}$ across roughly four orders of magnitude for a broad local galaxy sample, we examine how $M_{\star}$, $M_{\mathrm{BH}}$, and star formation rate behave across regimes. Because central galaxies anchor their own halos whereas satellites inhabit larger host halos, we focus on 57 centrals with complete measurements and find that $M_{\star}$ in star-forming galaxies and $M_{\mathrm{BH}}$ in quiescent galaxies follow a common power-law trend ($b \approx 1.13$) separated by a $d \approx 2.68\,\mathrm{dex}$ ($\sim 500\times$) offset, reflecting differences in overall efficiency of converting mass into a baryon-lifting effect. Similar trends hold across the wider sample beyond centrals. These shared scaling trends point to source-agnostic energetic self-regulation as stellar feedback yields to AGN feedback, producing the high-mass ``knee'' of the stellar-to-halo mass relation, and offer a unified transition framework and simulation benchmark with statistical performance comparable to separate fits.

\end{abstract}

\section{Introduction}
\label{intro}

Scaling relations provide a crucial window into the broad physical processes governing galaxy evolution, distilling the richness of baryonic physics into simpler, intuition-building targets for galaxy formation models to reproduce \citep[e.g.,][]{Somerville2015, Wechsler2018}. A chief example is the relationship between galaxy stellar mass and its dark matter (DM) halo mass --- the stellar mass--halo mass (SMHM) relation \citep{Moster2010, Behroozi2010, Wechsler2018}. The SMHM relation applies to `central' galaxies that are the dominant ones in their dark matter halos; `satellite' galaxies may depart from this relation if they have experienced significant tidal stripping and other environmental processing \citep{Wetzel2013,Behroozi2013}.  The SMHM relation rises steeply up to a characteristic break at $M_{\mathrm{halo}} \sim 10^{12}\,M_{\odot}$ ($M_{\star} \sim 3\times 10^{10}$--$10^{11}\,M_{\odot}$, corresponding to $L^{\star}$ galaxies). Above this mass, the SMHM flattens substantially, and most galaxies lack large amounts of recent star formation \citep[e.g.,][]{Kauffmann2003,Peng2010}. Because feedback from star formation alone cannot regulate gas cooling in these massive halos, cosmological models invoke feedback from accretion onto supermassive black holes (BHs), also known as active galactic nucleus (AGN) feedback, as the dominant regulator in this high-mass regime \citep[e.g.,][]{Bower2006, Croton2006} --- a regulation mechanism that cannot be captured using the SMHM relation alone. The goal of this paper is to rectify this limitation, giving a more coherent picture of the relative roles of stellar and BH feedback as a function of halo mass.

In the same way that the SMHM relation diagnoses galaxy evolution regulated by stellar feedback, scalings between BH mass and halo mass offer a direct probe of regulation driven by AGN feedback. Constraining a relation between two invisible massive components across vastly different spatial scales has understandably proved challenging, requiring dynamical measurements that resolve BH spheres of influence (tens to hundreds of parsecs) alongside large-scale ($\gg$\,kpc scale) tracers of the dark matter halo \citep[e.g.,][]{Ferrarese2002, Bandara2009, Marasco2021}. Despite these observational hurdles, $M_{\mathrm{BH}}$ is now well established to correlate with $M_{\mathrm{halo}}$ in massive galaxies, alongside galaxy-scale measures such as stellar velocity dispersion and bulge mass \citep{Gultekin2009,Kormendy&Ho2013}.

A question that might naturally emerge is whether there are similarities between how stellar feedback and AGN feedback interact with a central galaxy's interstellar medium (ISM) and circumgalactic medium (CGM). An extremely simplified framework that may help to explore this idea is the hypothesis of baryon lifting. This is explored most extensively in the context of AGN feedback. The basic idea is that gas cooling and accretion onto a central BH continue (or regulate) until the (cumulative) AGN feedback energy becomes comparable to the binding energy of the gas to the galaxy (and/or halo; e.g., \citealt{SilkRees1998}, \citealt{BoothSchaye2010}, \citealt{KingPounds2015}, \citealt{Voit2024}). Since the binding energy of gas to a galaxy and its halo should roughly scale as     $ G\frac{M_\text{b}M_{\mathrm{halo}}}{r} \sim G\frac{f_\text{b}M^2_\text{halo}}{r} \propto M_{\mathrm{halo}}^{5/3}$, and assuming that feedback energy is proportional to the (in this case) BH mass $M_{\mathrm{BH}}$, one might expect a scaling in the neighborhood of $M_{\mathrm{BH}} \propto M_{\mathrm{halo}}^{5/3}$, where differences from the predicted slope of $5/3$ would signal interesting physical processes, e.g., systematic trends in total mass profile (expected to be steeper for lower mass halos; \citealt{Dutton2014}), the fraction of gas that must be affected to regulate AGN feedback (expected to be larger for lower mass halos; \citealt{Oppenheimer2008}), or systematic trends in feedback efficiency \citep{BoothSchaye2010}. Such a BH mass--halo mass (BHHM) relationship is observed \citep{Marasco2021,Voit2024}.

In this picture, stellar feedback --- for central star-forming galaxies --- would scale with halo mass in similar ways. In this case, one might analogously expect the stellar feedback energy to be proportional to the stellar mass $M_{\star}$, and so $M_{\star} \propto M_{\mathrm{halo}}^{5/3}$ for star-forming galaxies. If one assumes $v_{\mathrm{max}} \propto v_{\mathrm{vir}}$, and noting that $M_{\mathrm{halo}} \propto v^3_{\mathrm{vir}}$, this maps directly onto the stellar mass Tully-Fisher relation \citep{BdJ2001,Reyes2011}, albeit with a slope slightly steeper than observed (fiducial slope $M_{\star} \propto v_{\mathrm{max}}^5$ vs.\ observed slope $M_{\star} \propto v_{\mathrm{max}}^{3.5-4.5}$). Again, the expectation of a slope $\sim5/3$ depends on how halo mass proxies vary with halo mass, and mass-dependent systematic trends in the distribution or state of matter and gas in galaxies, or the effectiveness of feedback \citep{DekelSilk1986,IKEA2020,Donahue&Voit2022}. The expectation would be that star-forming galaxies would adhere to a stellar mass--halo mass scaling (and their BHs could have a wide variety of masses). In contrast, under the ansatz that quiescence reflects domination by AGN feedback, quiescent galaxies would have a BH mass--halo mass scaling (where because stellar feedback is not dominant, stellar mass could scatter or scale differently). Crucially, this ansatz holds strictly only for central galaxies; satellite galaxies can be quenched externally by environmental processes (such as ram-pressure stripping in groups and clusters; \citealt{Wetzel2013}) without any connection to central BH feedback.

Qualitatively, the two separate SMHM and BHHM relations appear encouragingly consistent with this view. Yet, there are critical limitations that undermine the promise of a joint feedback-driven scaling between halo mass and the magnitude of feedback energy as a constraint on galaxy evolution theory. The SMHM relation is often built with large samples from cosmological surveys, with halo masses determined from number density \citep[e.g.,][]{Behroozi2010, Moster2010}, clustering, or weak lensing arguments \citep[e.g.,][]{Leauthaud2012, Zu2015}, although measures from individual group masses are consistent with these relations \citep[e.g.,][]{Yang2008, Lim2017}. In contrast, because BH mass is so challenging to measure, the BHHM relation is typically built from galaxies with directly-determined BH masses, using a wide variety of halo mass proxies, each with their own limitations and biases (e.g., group X-ray temperature, \citealt{Voit2024} --- observable practically only for relatively large halo masses \citealt{Bregman2018} --- galaxy velocity scales \citealt{Ferrarese2002}, or combinations of these measurements \citealt{Marasco2021}). Without exploring the joint scaling of stellar mass, BH mass, and halo mass using a common halo mass proxy, it is difficult to understand if apparent differences between the two scaling relations reflect systematic biases among halo mass proxies, or genuine differences in how stellar and AGN feedback operate.

Our goal in this paper is to circumvent this challenge by adopting a single proxy for halo mass that is validated across the whole range of halo masses relevant to the transition from stellar feedback-dominated to AGN feedback-dominated: globular cluster system masses.  \citet{Hudson2014} demonstrate a linear relationship between $M_\text{GCS}$, the total integrated mass of a galaxy's globular cluster system, and $M_{\mathrm{halo}}$, as determined from stellar masses calibrated by weak lensing. This relation appears to be approximately linear in slope \citep{Dornan2023,Dornan2025,Dornan2026}, and applies over the halo mass range $10^{10}M_\odot \lesssim M_{\mathrm{halo}} \lesssim 10^{14}M_\odot$. While the origin of the approximate linearity of this relationship is debated \citep{Chen2024,Forbes2018,Harris2017}, the GC mass--halo mass scaling provides a single tool that can be uniformly applied across a factor of $10^4$ in halo mass to investigate how stellar mass, BH mass, and star formation activity vary with halo mass, across the critical halo mass of $\sim 10^{12}M_\odot$ that marks the transition from stellar feedback-dominated to AGN feedback-dominated. 

The remainder of the paper is structured as follows. Section \ref{data} discusses the estimates of $M_\text{BH}$, $M_{\star}$, $M_{\mathrm{halo}}$, and SFR that we have adopted, where our main constraint on the sample is the availability of GC system measurements with which to estimate the $M_{\mathrm{halo}}$. We discuss also our identification of a central galaxy subsample in our dataset. Section \ref{results} presents our derived scalings of $M_{\star}$ and $M_\text{BH}$ with $M_{\mathrm{halo}}$ for star-forming and quiescent central galaxies, and investigates the possibility of a joint relationship between the four quantities that describes how feedback energy --- independent of origin --- scales with halo mass. In Section \ref{discussion}, we discuss some implications of and limitations in our understanding of a common origin-independent relation between feedback energy and halo mass, in addition to exploring how satellite galaxies populate this relation, and in Section \ref{summary} we present a brief summary.

\section{Data}
\label{data}

\subsection{Halo Mass}

One of the important aspects of this work is our adoption of globular cluster (GC) system masses as a uniform method for halo mass estimation, applicable across a wide range of halo masses \citep[see also][]{Spitler2009, Peng2016}. In this section, we describe our GC system-derived halo masses, and compare them with independent values of halo mass for the systems in our sample.  

\subsubsection{Halo Masses from Globular Clusters} \label{gcs-halo}

To obtain halo mass estimates across a wide mass range using uniform methodology, we use the near-linear scaling between total globular cluster system mass ($M_{\mathrm{GCS}}$) and dark matter halo mass ($M_{\mathrm{halo}}$). Following \citet{Hudson2014}, we adopt the conversion
\begin{equation}
    \eta = \frac{M_{\mathrm{GCS}}}{M_{\mathrm{halo}}} = (3.9 \pm 0.9) \times 10^{-5},
\end{equation}
and apply it to the 418 galaxies with $M_{\mathrm{GCS}}$ measurements in the \citet{Harris2013} catalog, accounting for the reported $\sim 0.2\,\mathrm{dex}$ intrinsic scatter in the relation. Empirically, $M_{\mathrm{GCS}}$ tracks halo mass roughly linearly across a broad mass range, making it a useful halo-mass indicator in regimes where hot X-ray gas is faint or kinematic tracers do not extend to large radii \citep[e.g.,][]{Spitler2009, Hudson2014, Harris2017, Burkert2020}. Although globular cluster systems reflect a combination of \textit{in situ} formation and hierarchical accretion, and the physical drivers behind the near-linearity of the $M_{\mathrm{GCS}}$--$M_{\mathrm{halo}}$ relation remain active topics of study even as models often recover roughly linear trends \citep[e.g.,][]{Harris2017, Forbes2018, Chen2024}, we treat it here primarily as an empirical scaling, discussing the implications of its limitations as a halo mass indicator later in the paper.

We note that our calibration of $\eta$ is not  independent of prior stellar-to-halo mass relations. \citet{Hudson2014} tie their calibration to the $z=0$ weak-lensing-based stellar-to-halo mass relation of \citet{Hudson2015} (built on CFHTLenS and the \citealt{Moster2010} parameterization), an approach widely used in subsequent calibrations of the $M_{\mathrm{GCS}}$--$M_{\mathrm{halo}}$ scaling \citep[e.g.,][]{Spitler2009, Harris2013, Harris2017, Dornan2023, Dornan2025, Dornan2026}. Independent checks using dynamical tracers, mass-to-light ratios, or fundamental-plane relations \citep[e.g.,][]{Georgiev2010, Forbes2018, Zaritsky2022} broadly support near-linearity from dwarfs (where mean GC counts can fall below unity; \citealt{Forbes2018, Dornan2026}) up to massive clusters, with post-2015 estimates of $\eta$ clustering around $(2.5\text{--}4.0) \times 10^{-5}$ depending on GC count/mass definitions, halo mass conventions, and sample selection \citep{Dornan2023}. Even if an alternate $\eta$ within this range were adopted, individual $M_{\mathrm{halo}}$ estimates would shift by at most $\sim 0.2\,\mathrm{dex}$, and a global offset in $\eta$ would leave our best-fit power-law slopes and relative offsets unchanged.

While earlier studies have used $M_{\mathrm{GCS}}$ to infer halo masses for well-studied individual systems or targeted subsets \citep[e.g.,][]{Spitler2009, Peng2016}, applying this proxy across a broad sample allows us to examine joint trends in inferred halo mass with stellar and black hole mass using a single halo-mass indicator. Nevertheless, the \citet{Harris2013} compilation is heterogeneous: it aggregates measurements from many independent studies with varying area coverages, background subtraction strategies, and completeness corrections. In practice, literature GC catalogs skew toward more luminous, massive, and early-type/quiescent galaxies (as well as nearby systems) whose rich GC populations are more readily detected and extrapolated than those in low-mass or star-forming systems. We therefore treat our sample as a targeted, proxy-enabled compilation rather than a volume-complete census, and next evaluate its systematics directly against other halo-mass estimators.

\subsubsection{Comparisons with Other Literature Halo Masses}

\begin{figure}
    \centering
    \includegraphics[width=\linewidth]{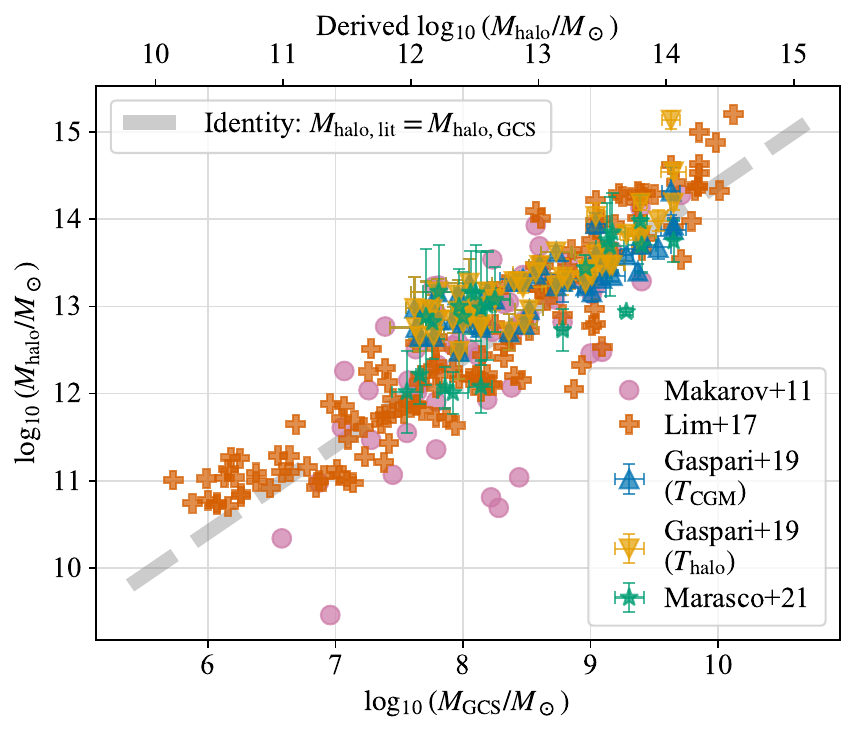}
    \caption{Comparison of several literature $M_{\mathrm{halo}}$ measurements. The bottom horizontal axis is the base 10 logarithm of the globular cluster system mass ($M_\text{GCS}$) from \citet{Harris2013}. The top horizontal axis uses the transformation $M_{\mathrm{halo}} = M_\text{GCS}/\eta$ \citep{Hudson2014} to derive halo mass from globular cluster system mass. The vertical axis represents the halo mass estimate from one of the following: Pink circles and orange crosses are the group-based halo masses from \citet{Makarov2011} and \citet{Lim2017}, respectively, both of which lack errorbars in the source publication. We interpret the group mass as the halo mass of the associated BCG. Blue triangles and yellow inverted triangles are X-ray temperature-derived halo masses from \citet{Gaspari2019}, using the temperature-to-mass conversion from \citet{Voit2024}. Green stars show the compiled halo masses of \citet{Marasco2021}, which use inhomogeneous methods. The light grey dashed line corresponds to points where the derived halo mass from globular cluster system mass is equal to the halo mass estimate from the other source.}
    \label{fig:halo_mass_comparison}
\end{figure}

The invisibility of dark matter compels us to compare our $M_{\mathrm{halo}}$ values against several other catalogs of halo mass estimates in the literature. Figure \ref{fig:halo_mass_comparison}  summarizes this comparison showing broad agreement in aggregate alongside systematic offsets and tilts for individual proxies. \cite{Marasco2021} use a highly inhomogeneous sample derived from rotation curves, globular cluster dynamics, Schwarzschild models, and X-ray gas modeling. The two \cite{Gaspari2019} halo mass estimates are calculated using equation (6) of \cite{Voit2024} to convert the $T_\text{CGM}$ and $T_\text{halo}$ measurements from the former paper to $M_{\mathrm{halo}}$. \cite{Makarov2011} and \cite{Lim2017} both report group-based masses, with the former using a projected kinematic group mass \citep{Heisler1985} and the latter using an iterative abundance matching technique based on \cite{Lu2016} and \cite{Yang2005}. Since both of these studies report one mass value per group, we set that group mass to be the halo mass of the associated central galaxy for comparison. The two group-based $M_{\mathrm{halo}}$ estimates probe the majority of the dynamic range of our sample, while the \cite{Marasco2021} and \cite{Gaspari2019} samples are biased to higher masses, as SMBH scaling relations were a primary focus of their analyses. 

Across all compared catalogs, the median offset $\Delta \log M_{\mathrm{halo}} \equiv \log(M_{\mathrm{halo,lit}}/M_{\mathrm{halo,GCS}})$ is 0.14 dex, with 16th--84th percentile intervals of -0.28 to 0.56 dex. Even so, individual samples exhibit non-unity slopes relative to our $M_{\mathrm{GCS}}$-based estimates, and pairwise tilts are present across all proxy pairs examined. We examine how these systematic differences affect the inferred power-law slopes and model benchmarks in Section~\ref{slope talk}.

\subsection{Black Hole Mass}
\label{black holes}
Following \citet{Gultekin2009}, we restrict our black hole mass compilation to dynamical measurements to avoid systematic biases introduced by secondary or scaling-calibrated estimators. The first sample we consider is that of \cite{Terrazas2017}, which is itself built off of the $M_\text{BH}$ estimations compiled by \cite{Saglia2016} and \cite{vdB2106}. We augment these 91 $M_\text{BH}$ estimates with 13 found in \cite{Marasco2021} (which are sourced from \cite{Saglia2016}, \cite{Kormendy&Ho2013}, and \cite{Kranovic2018}), leaving us a sample of 104 galaxies with robust, dynamical $M_\text{BH}$ estimates and diverse morphological types. While this compilation draws together 104 dynamical $M_{\mathrm{BH}}$ measurements across diverse morphologies, resolving the black hole sphere of influence favors nearby, bright, or massive systems—so the sample reflects observational accessibility rather than an unbiased census of the local galaxy population.

\subsection{Stellar Mass}
\label{stellar mass}
Following \cite{Terrazas2017} and \cite{Marasco2021} we derive stellar mass estimates using the 2MASS K-band luminosity \citep{Huchra2012} and the mass to light ratio $M_{\star}/L_\text{K} = 0.6$ suggested in \cite{Schombert&McGaugh}, assuming a uniform uncertainty of 0.15 dex for all values of $M_{\star}$. This ratio is sometimes reported to be as high as $M_{\star}/L_\text{K} = 0.75$ as it was in \cite{Terrazas2017}, but the choice of conversion factor has negligible impact on our results.

Comparison with the \cite{Leroy2019} catalog, which derive their stellar masses from the SED fits of \cite{Salim2016}, shows excellent agreement. The 16th, 50th, and 84th percentiles of $\Delta\log M_* \equiv \log (M_{*,\text{Leroy+19}} - M_{*,\text{K-band}})$ are -0.093 dex, 0.007 dex, and 0.101 dex respectively.

\subsection{Star Formation Rate}
\label{sfr}

Given that the sole function of star formation rate (SFR) in our analysis is to establish a binary quiescence flag, we prioritize coverage over homogeneity in constructing our sample. We start with the SFR values calculated by \cite{Leroy2019}, which we use for 207 systems. We then proceed with several photometric SFR estimations widely used in the literature.

Following \cite{Terrazas2016}, we calculate star formation rates with IRAS far-infrared fluxes (\cite{Rice1988}, NED corrections to \cite{Knapp1989}, \cite{Moshir1990}, \cite{Sanders2003}). First, we get the far-infrared (FIR) flux using equation (A1) in \cite{Bell2003}.

The same paragraph tells us that the total infrared (TIR) flux is simply $\mathrm{TIR} = 2 \times \mathrm{FIR}$.
We can then use equation (12) (modified from ergs to watts) of \cite{Kennicutt2012} to calculate the SFR

\begin{equation}
\log_{10}\mathrm{SFR_{TIR}}(M_\odot\;\mathrm{yr}^{-1}) = \log_{10}(L_\mathrm{TIR})-36.41.
\end{equation}

Our next SFR source are 70 micron fluxes from Spitzer/MIPS. We pull values first from the Spitzer Local Volume Survey \citep{Dale2009}, then from the Spitzer Observations of Passive and Star-Forming Early-Type Galaxies \citep{Temi2009}. We convert the 70 $\mu$m fluxes from these sources to SFRs via equation 22 of \cite{Calzetti2010}.

Our final SFR source are the 12 micron (``W3'') and 23 micron (``W4'') fluxes from AllWISE \citep{Cutri2013}. We calculate the resulting SFRs using equations 1 and 4 from \cite{Cluver2017}.

For all undetected sources, we then consider SFR upper limits for all of the above sources in the same order as presented above.

We use these SFR estimates in conjunction with the stellar masses of Section~\ref{stellar mass} to compute $\mathrm{sSFR} \equiv \mathrm{SFR}/M_*$. Following common convention \citep[e.g.,][]{Wetzel2013}, quiescent galaxies are those with $\mathrm{sSFR} < 10^{-11}\,\mathrm{yr^{-1}}$, and $\mathrm{sSFR} > 10^{-11}\,\mathrm{yr^{-1}}$ for star-forming galaxies. Notably, some famous spiral galaxies with measured $M_\text{BH}$ (e.g., M31, M81) are marked as quiescent by our process.

\subsection{Centrality Flag}
\label{central method}
In order to identify central galaxies in our sample, we use the 2MASS Redshift Survey (2MRS) group catalog of \cite{Lim2017}. Their group catalog (\texttt{2MRS+(L) group.dat}) gives the 2MASS ID of the central (in their methodology, central $\equiv$ most luminous) galaxy of each group. This allows us to select only the central galaxies from their galaxy catalog (\texttt{2MRS+(L) galaxy.dat}). Matching our sample against the central galaxy subsample of \cite{Lim2017} provides us with a flag for identifying central galaxies. It should be noted that the absence of this flag does not necessarily imply that a given galaxy is a satellite, as it could simply be absent from \texttt{2MRS+(L) galaxy.dat}. This method positively identifies 143 central galaxies in the \cite{Harris2013} catalog.

\subsection{Sample Constraints and Intersections}
We draw our primary analysis sample of 57 local central galaxies as follows. We start with the \cite{Harris2013} Catalog of Globular Cluster Systems in Galaxies. Of the 422 entries in that catalog, 418 have usable $M_\text{GCS}$ values which we convert to $M_\text{halo}$ in the process described in Section~\ref{gcs-halo}. From there, sSFRs are needed to identify systems as star-forming or quiescent. The requirement of SFRs (\S \ref{sfr}) brings our sample to 354, which the requirement of $M_*$ (\S \ref{stellar mass}) reduces further to 325. 270 of these galaxies are present in the \cite{Lim2017} catalog (\S \ref{central method}), 143 of which are identified as centrals. 

Selecting those galaxies with $\log(\mathrm{sSFR}) > -11$ yields 32 star-forming central galaxies for which we have stellar masses. Selecting for $\log(\mathrm{sSFR}) < -11$ and intersection with our black hole sample (\S \ref{black holes}) yields 25 quiescent central galaxies with $M_\text{BH}$ estimates.
These 57 galaxies form the sample of centrals with the necessary measurements (stellar masses for star-forming centrals, with the addition of BH masses for quiescent centrals) for our analysis.

In addition, we analyze also the full sample, irrespective of status as central, satellite or unknown, with required measurements for our scaling relations --- 73 star-forming galaxies with stellar masses, and 40 quiescent galaxies that also have a black hole mass, for a total of 113 galaxies. 57 of these are our core sample of centrals, 27 are in \texttt{2MRS+(L) galaxy.dat} but not identified as centrals (satellites), and 29 are absent from the \cite{Lim2017} catalog and thus have unknown centrality status.

\section{Analysis and Results}
\label{results}

\subsection{Model Construction}
\label{model}

The literature presents us with two feedback proxy scaling relations, $M_*-M_\text{halo}$ and $M_\text{BH}-M_\text{halo}$. Restricting these relations to the sSFR regimes where their feedback modes are most dominant, their combined description requires six free parameters: a slope, intercept, and intrinsic scatter ($b$, $a$, $\sigma$) for both the star-forming SMHM relation and the quiescent BHHM relation. Our interpretation of the baryon lifting hypothesis suggests that a simple underlying mechanism (regulation via feedback with binding energy as a benchmark) permits fewer parameters. We therefore condense these two relations to a single power law with a single additional parameter ($d$) to account for the efficiency differences of these two distinct astrophysical processes. We then find the maximum likelihood fits of these two (six parameter, four parameter) models to our data and compare their effectiveness.

We choose to account for only the feedback proxy mass of the dominant feedback mode rather than a combined feedback energy proxy (e.g. $M_* + d\times M_\text{BH}$) for a few reasons. First of all, the number of primary measurements of $M_{\text{BH}}$ is small, and disproportionately samples high-mass, quiescent systems. Therefore, a combined feedback proxy would either have a much smaller sample size than our model, or set $M_\text{BH} = 0$ for the majority of systems, which would not accurately portray total cumulative feedback. Second, the energetic contribution from the non-dominant feedback mode is vanishingly small for most systems in our sample, owing to the large dynamic range covered by stellar masses and BH masses. A hypothetical system with equal stellar and AGN feedback contributions would be offset by $\log_{10}(2) \approx 0.3$ dex vertically from its position in our model, and the largest such offset in our sample in practice is only $\sim0.15$ dex. This is well within the intrinsic scatter of $\sigma \approx 0.46$ dex we measure in Section \ref{results}. The insignificance of summing stellar and AGN feedback contributions demonstrates a strong bimodality in the dominant feedback mode, which mitigates the conceptual discomfort of using an instantaneous flag (sSFR) to choose between two time-integrated energy proxies.

We acknowledge that this is not the only way we could have characterized the null hypothesis (for instance, we could have made a cut in $M_\text{halo}$ instead of sSFR to separate dominant feedback regimes). Likewise, we do not claim that the literature consensus is that the SMHM and BHHM relations indicate totally disparate phenomena. These models represent two simplistic extremes. Neither should be expected to thoroughly explain the rich complexity of any individual galactic system without an abundance of additional context.

\subsection{Results}

\begin{figure}[h]
\centering
\includegraphics[width=\linewidth]{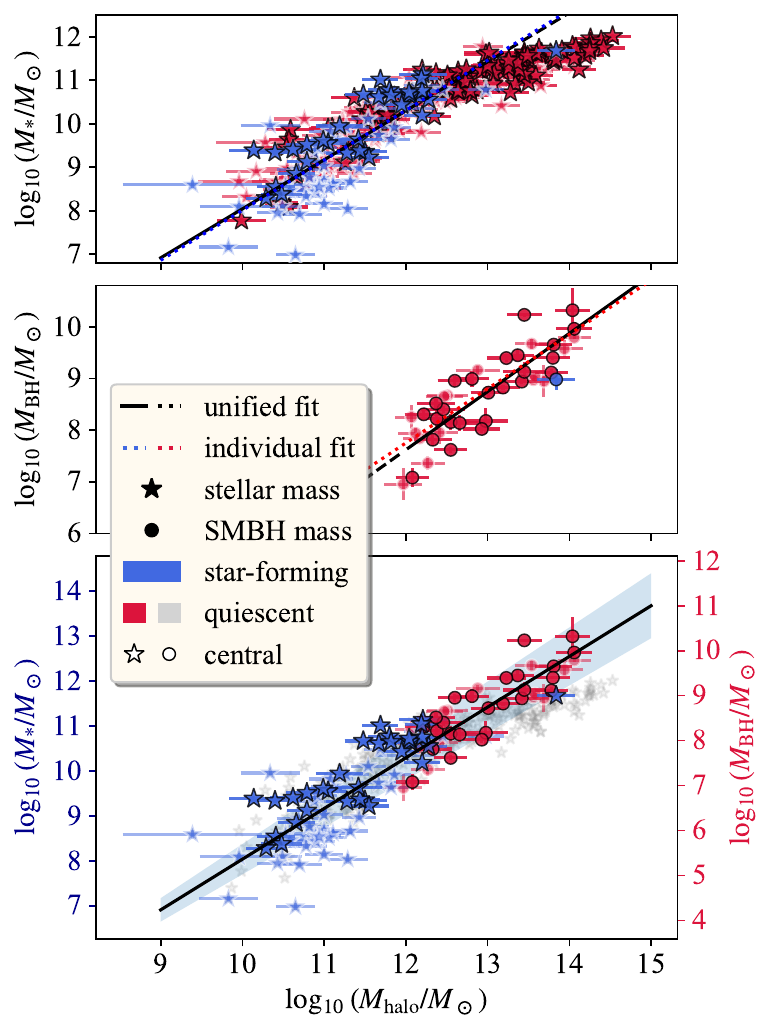}
\caption{A comparison of separate feedback proxy scaling relations against a unified feedback relation.
\\\textbf{Panel (a)} shows the SMHM ($M_*-M_\text{halo}$) relation in log-log space for 325 systems. A blue dotted line shows the best fit line for the star-forming central SMHM relation ($n=32$, blue stars with black outline). A solid black line shows the unified feedback relation through $M_\text{halo} \leq 10^{12} M_\odot$, and a dashed black line shows that relation for $M_\text{halo} > 10^{12} M_\odot$. 
\\\textbf{Panel (b)} shows the BHHM ($M_\text{BH}-M_\text{halo}$) relation for 42 systems. A red dotted line shows the best fit line for the quiescent central BHHM relation ($n=25$, red circles with black outline). A solid black line shows the unified feedback relation through $M_\text{halo} \geq 10^{12} M_\odot$, and a dashed black line shows that relation for $M_\text{halo} < 10^{12} M_\odot$. 
\\\textbf{Panel (c)} shows our unified feedback relation. All stellar masses (325 blue and grey stars) use the left vertical axis $\log_{10}(M_*/M_\odot)$. All black hole masses (41 red circles) use the right vertical axis $\log_{10}(M_\text{BH}/M_\odot)$, which is offset from the left vertical axis by $d = 2.68$. A solid black line shows the unified feedback relation for central galaxies ($n=57$, all shapes with black outlines), and a light blue shaded region shows the error in the fit.}
\label{combined}
\end{figure}

Panel (c) of Figure~\ref{combined} showcases our unified model wherein feedback energy regulates the atmospheres of central galaxies through baryon lifting, agnostic of the source of that energy. Stellar feedback performs the work for baryon lifting in star-forming galaxies, and AGN feedback does the same for quiescent ones. Thus, their mass-based feedback proxies populate the same power-law after accounting for (empirical, but anticipated) differences in effectiveness between the two modes. That power law follows the functional form:
\begin{align}
    (\log_{10}M_*)\; \texttt{or}\; &(\log_{10}M_\text{BH}+d) =\nonumber\\
    &a+b(\log_{10}M_\text{halo}- \log_{10}M_0) + \sigma, \label{power law}
\end{align}
where $M_*$ is selected when $\log({\mathrm{sSFR})} > -11$, and $M_\text{BH}$ is selected when $\log(\mathrm{sSFR}) < -11$, and $M_0$ is the mean log halo mass, $\langle\log(M_\text{halo})\rangle = 11.85$.

The best fit parameters for our central galaxy subset are as follows:

\begin{itemize}
    \item Overall intercept $a$: $10.126 \pm 0.054$
    \item Power law slope $b$: $1.127 \pm 0.054$
    \item Intercept offset $d$: $2.679 \pm 0.118$
    \item Intrinsic scatter $\sigma$: $0.464 \pm 0.032$ 
\end{itemize}

\begin{figure}[h]
\centering
\includegraphics[width=\linewidth]{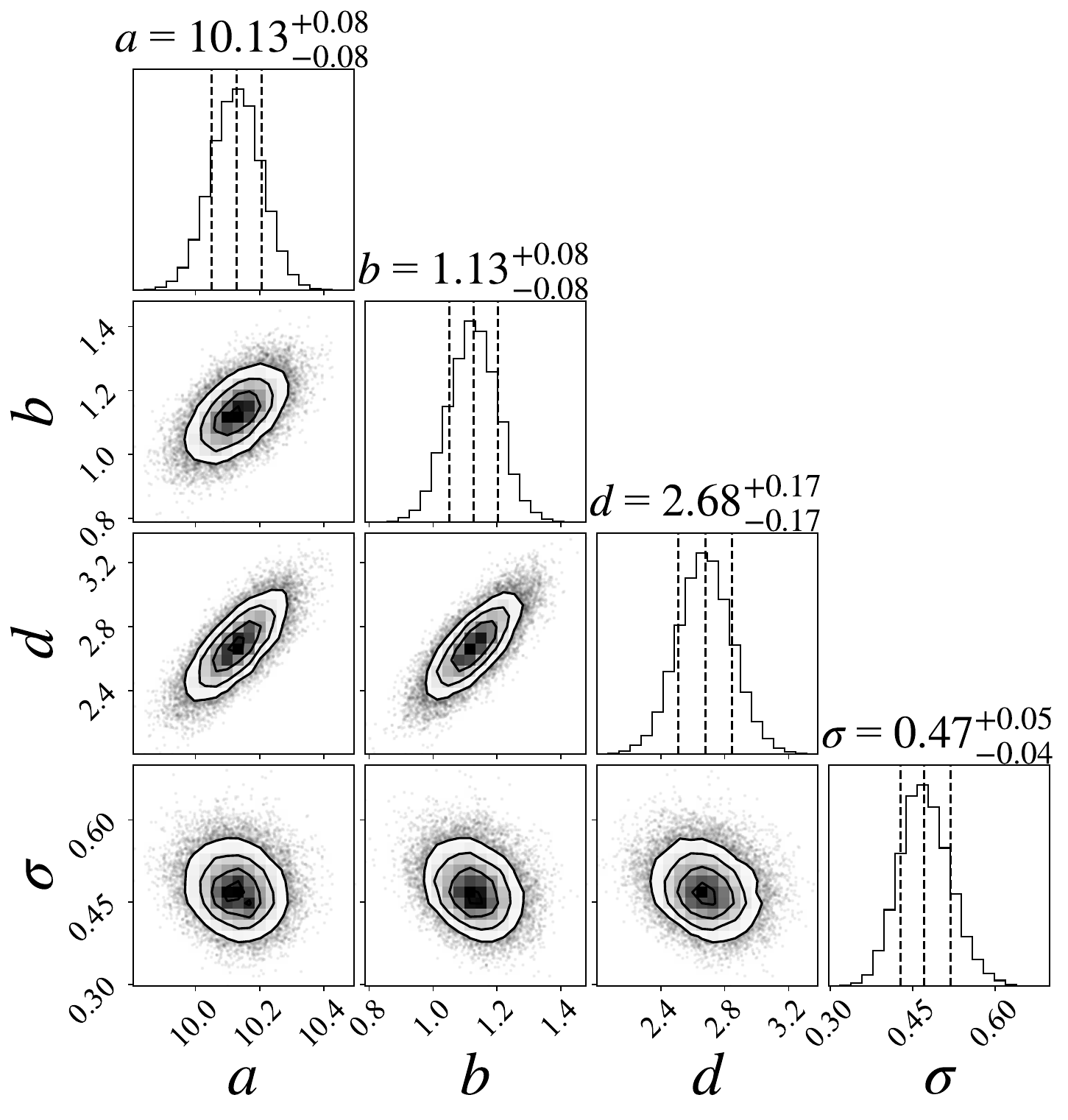}
\caption{Corner plot showing the distribution of parameters in the MCMC fit of Equation \ref{power law} for our central galaxy dataset. Histograms along the main diagonal show 1-D posterior distributions of the intercept ($a$), slope ($b$), offset $(d)$, and intrinsic scatter($\sigma$). Vertical dotted lines show the 16th, 50th, and 84th percentiles of each 1-D distribution. Non-diagonal plots show the 2-D joint posterior projections of the corresponding variables. Contour lines show the 11.8th, 39.3th, 67.5th, and 86.4th percentiles of each 2-D distribution.} 
\label{corner}
\end{figure}

Since stellar feedback does not play a substantial role in the energy balance for quiescent systems, the $M_{\star}$--$M_{\mathrm{halo}}$ relationship decouples from the power law and flattens out (grey stars in panel (c) of Figure \ref{combined}).

The slope of the power law fit $b = 1.13 \pm 0.05$ is on the low end of the range expected from the literature on scaling relations, bounded above by the total baryon binding energy scale $U_\text{b} \propto M_\text{halo}^{5/3}$. We elaborate on the physical implications of this slope (and limitations thereon) in Section~\ref{slope talk}. 

A corner plot for the above functional form given our dataset is provided in Figure~\ref{corner}. It reveals moderate pairwise correlation between the fit parameters $a$, $b$, and $d$, with the strongest effect between $b$ and $d$. The measured power-law slope is sensitive to differences in the measured feedback effectiveness offset and vice versa.

Perhaps surprisingly, the fitting procedure performed on the central galaxy subsample yielded nearly identical fit parameters to that of the full sample, with only the error on the intercept offset showing any difference ($d_\text{full sample} = 2.679 \pm 0.136$). 

To test whether the unified four-parameter model is statistically favored over the six-parameter baseline of separate star-forming ($M_{\star}$--$M_{\mathrm{halo}}$) and quiescent ($M_{\mathrm{BH}}$--$M_{\mathrm{halo}}$) power laws introduced in Section~\ref{model}, we compare their fits using the Bayesian Information Criterion (BIC). The BIC for a given model is calculated as follows:
\begin{equation}
    \mathrm{BIC} = k\ln(n) - 2\ln(\mathcal{L}),
\end{equation}
where $k$ is the number of parameters in the model, $n$ is the number of individual datapoints being fit in the sample, and $\mathcal{L}$ is the (maximized) likelihood of that model \citep{Schwarz1978}. When comparing the performance of two models with the BIC, $n$ should be the same in both cases, reflecting the fact that the data are unchanged. Our unified relation makes use of 4 parameters ($a$, $b$, $d$, and $\sigma$), whereas the model which treats the $M_*$--$M_{\text{halo}}$ and $M_{\text{BH}}$--$M_{\text{halo}}$ separately fits an intercept, slope, and intrinsic scatter for each relation for a total of 6. The $\Delta$BIC between the two models was found to be 3.46 in favor of the unified model, which suggests a moderate statistical favorability \citep{Liddle2004}.

The Akaike Information Criterion (AIC) is another common metric for selecting between models whose varying complexity lies in their parameterization. The AIC is calculated as follows:
\begin{equation}
    \mathrm{AIC} = 2k - 2\ln(\mathcal{L}),
\end{equation}
with variables having the same meaning as in the above BIC definition \citep{Akaike1974}. The $\Delta$AIC between the two models was found to be 1.99 in favor of the naïve model, which means it is slightly favored over the unified model \citep{Akaike1974, Liddle2004}. Neither the AIC nor BIC results are particularly conclusive either way. The BIC imposes a greater penalty on additional parameters than the AIC, given that $\ln(n) > 2$ for integers $n \geq 8$.

\section{Discussion}
\label{discussion}

The baryon lifting picture suggests that the main action of feedback is to reshape the circumgalactic medium of a galaxy, regulating the amount of feedback needed to maintain a galaxy in such a state.  One of the expectations of such a picture is that galaxies might be expected to follow a joint relationship between halo mass and tracers of the two main feedback channels --- either stellar mass or BH mass --- with a common slope but an offset indicating the difference in efficiency between the two modes of feedback. We have tested this paradigm using galaxies with globular cluster system mass measurements, which scale roughly linearly with halo mass for both star-forming and quiescent galaxies across a four-order-of-magnitude range in halo mass. We find that a joint fit to the $M_*$--$M_{\text{BH}}$--$M_{\text{halo}}$ relation, with power-law slope $b\approx1.1$ and $\log_{10}$ efficiency offset $d\approx2.7$ has comparable statistical quality to separate regime-specific fits, while offering a unified framing of the star-forming-to-quiescent transition. In the following sections, we examine what this unified scaling implies for the high-mass SMHM turnover and outlier galaxies (Section~\ref{knee}), evaluate how proxy systematics and cooling physics shape the inferred slope and offset (Section~\ref{slope talk}), and consider how satellite galaxies fit into this picture (Section~\ref{central discussion}).

\subsection{The `knee' of the SMHM relation and the emergence of quiescent galaxies}\label{knee}

One immediate implication is that the `knee' at high halo masses in the SMHM relation is an artifact of considering how stellar mass scales with halo mass, in a regime where stellar mass is now decoupled from the dominant feedback mechanism that regulates gas cooling \citep[see also][]{Mandelbaum2016, Posti2019, Donahue&Voit2022}. When instead $M_{\star}$ is considered for star-forming galaxies, and $M_{\mathrm{BH}}$ is considered for quiescent, AGN feedback-dominated galaxies --- scaled by an empirical offset reflecting differences in feedback efficiency --- these quantities (a stand in for feedback energy) follow an unbroken scaling law across four orders of magnitude in halo mass. Physically, this suggests that as a star-forming galaxy grows its supermassive black hole, its feedback eventually exceeds that available from star formation \citep[e.g.,][]{Bower2017, Terrazas2020, Bluck2020}, quenching the galaxy and placing it onto a BH mass--halo mass correlation irrespective of its stellar mass. In this picture, the high-mass end of the SMHM relation then reflects the combination of the stellar mass that a galaxy had when it became AGN feedback-dominated \citep{Peng2010}, growth by any residual (energetically unimportant) star formation, and growth through galaxy merging \citep{Oser2010}.

In this picture, the growth of a galaxy's supermassive black hole is of central importance in determining its evolutionary fate. Around $M_{\mathrm{halo}} \sim 10^{12}\,M_\odot$, supernova-driven outflows lose buoyancy against the deepening gravitational potential and emerging hot circumgalactic medium \citep[e.g.,][]{Keller2016, Bower2017, Henriques2019}. As stellar feedback becomes ineffective at lifting gas out of a galaxy, cooling gas accumulates, driving angular momentum transport and fueling rapid growth of the central SMBH. If SMBH growth were suppressed, such unchecked gas cooling in massive halos would produce hyper-massive, actively star-forming central galaxies --- a common outcome of simulations that lack AGN feedback \citep[e.g.,][]{White1991, Benson2003, Croton2006}. Instead, central SMBH growth appears to be the norm in massive halos, and AGN feedback, being orders of magnitude more energetic per unit mass than stellar feedback, is sufficiently powerful to heat and reshape the ISM and CGM, allowing the galaxy to become quiescent.

An exception that illustrates and enriches this picture is the population of massive star-forming ``super spirals'' \citep{Bogdan2013, Ogle2016, Ogle2019, Posti2019, Ray2024}, disk galaxies that reside in halos of $M_{\mathrm{halo}} \sim 10^{13}\,M_\odot$ yet remain on the extrapolated star-forming $M_{\star}$--$M_{\mathrm{halo}}$ sequence. In our framework, these systems represent rare massive halos that somehow avoided rapid central SMBH growth (for instance, due to high angular momentum or a quiet merger history; \citealt{Kormendy&Ho2013, Posti2019, DiTeodoro2023}), forcing ongoing star formation to continue carrying the halo's regulatory feedback load despite its far lower efficacy. This tension manifests in galaxies with undermassive stellar populations for their halos \citep{Ogle2019} and slowing star formation rates \citep{Ray2024} as they deplete their baryon reservoirs \citep{Posti2019}.

While not spirals morphologically, two galaxies in our sample occupy the same high-mass, high-sSFR corner of the parameter space, and appear to play a conceptually similar role. The two star-forming systems in Figure~\ref{combined}(a) with $M_\text{halo} \gtrsim 10^{13}M_\odot$ are Centaurus A (NGC 5128) and Perseus A (NGC 1275), in order of increasing halo mass. Both of these galaxies' stellar masses fall below the unified feedback relation line. Panel (b) shows that Perseus A's SMBH is also undermassive relative to the relation. It was not included with our sample, but \cite{Cappellari2009} report Centaurus A has $\log(M_\text{BH}/M_\odot) \approx 7.74$, which is likewise undermassive. Within the baryon lifting framework, these systems could both be considered ``dysregulated'', with insufficient AGN feedback perpetuating continued star formation. Notably, both systems' SMBHs are active and accreting \citep{Israel1998, Morganti2023}. Given that $M_*$ and $M_\text{BH}$ trace cumulative feedback energy over cosmic history, it is plausible that these systems are actively moving toward a more well-regulated state by growing their SMBHs.

\subsection{Interpreting our joint fit parameters in the baryon lifting paradigm}
\label{slope talk}

Interpreting the empirical fit parameters in Figure~\ref{combined} within the baryon lifting framework and translating them into useful constraints for galaxy formation models requires separating what is robust from what is sensitive to methodology. Specifically, our joint fit yields two key quantities: the global power-law slope ($b = 1.13 \pm 0.05$) and the vertical offset between the star-forming $M_{\star}$ and quiescent $M_{\mathrm{BH}}$ sequences ($d \approx 2.68\text{ dex}$, corresponding to a factor of $\sim 500$).   

Regarding the slope, observational halo mass systematics play an important role in how tightly $b$ should be interpreted. While $M_{\mathrm{GCS}}$-derived halo masses agree in aggregate with other methods, Figure~\ref{fig:halo_mass_comparison} shows that important pairwise systematic tilts are typical across all halo mass proxies, directly impacting the inferred power-law slope. For example, \citet{Voit2024} analyze the $M_{\mathrm{BH}}$--$M_{\mathrm{halo}}$ relation using X-ray temperatures from \citet{Gaspari2019}, finding $M_{\mathrm{BH}} \propto M_{\mathrm{halo}}^{1.6}$ when using $T_{\mathrm{CGM}}$ and $M_{\mathrm{BH}} \propto M_{\mathrm{halo}}^{1.3}$ when using $T_{\mathrm{halo}}$. Because both X-ray proxies tend to assign higher halo masses ($\sim 10^{13} M_\odot$) for systems with GC-derived values $\sim 10^{12} M_\odot$ (blue and yellow triangles in  Figure~\ref{fig:halo_mass_comparison}), they compress the horizontal dynamic range and steepen the slope by $\Delta b \approx 0.2$--$0.5$ relative to our $M_{\mathrm{GCS}}$-based fit. 
Likewise, while empirical calibrations find the $M_{\mathrm{GCS}}$--$M_{\mathrm{halo}}$ relation to be approximately linear across four orders of magnitude \citep[e.g.,][]{Hudson2014, Harris2017, Dornan2023}, any modest departure from exact linearity, or differing slope/scatter at the low- or high-mass ends \citep[e.g.,][]{Forbes2018, Bastian2020, Chen2024}, would gently stretch or compress the inferred halo-masses and shift the recovered slope at roughly the $\Delta b\sim 0.1$ level. Combined with $\eta$ zero-point uncertainties ($\sim 0.2\,\mathrm{dex}$) and the moderate covariance between $b$ and the feedback efficiency offset $d$, this reinforces treating $b = 1.13 \pm 0.05$ as an indicative benchmark ($b \approx 1.1$, with tolerances of 0.1--0.2 in slope) rather than a rigid value accurate to the second decimal place.

Beyond these observational systematics, interpreting $b \approx 1.13$ relative to the naive $5/3$ virial expectation raises a conceptual question: does a virial mass $M_{\mathrm{halo}}$ even represent the total mass relevant to baryon lifting arguments across all mass regimes? In massive halos ($M_{\mathrm{halo}} \gtrsim 10^{13}\,M_\odot$), outer CGM gas has cooling times comparable to or exceeding the Hubble time \citep{Donahue&Voit2022} and is effectively decoupled from the baryon cycle. If feedback only regulates gas within a cooling radius that encloses a smaller fraction of the virial mass in larger halos, evaluating feedback against $M_{\mathrm{halo}}$ naturally flattens the observed slope below the naive $5/3$ virial scaling. Conversely, at low halo masses, cooling times well below $1\text{ Gyr}$ imply repeated galactic-fountain recycling, meaning that the gas binding energy may understate the cumulative energetic load required to regulate star formation. Spanning four orders of magnitude in $M_{\mathrm{halo}}$ simultaneously exposes these mass-dependent variations in $\log t_{\mathrm{cool}}$ in a way that narrower-range SMHM or BHHM studies might not.

Turning to the offset, the $\sim 2.68\text{ dex}$ separation between star-forming $M_{\star}$ and quiescent $M_{\mathrm{BH}}$ offers a more direct target for cosmological simulations to reproduce \citep[see][for a review]{Somerville2015}. Within the baryon lifting paradigm, if populations dominated by stellar and AGN feedback both regulate the same fraction of the halo binding energy at a given $M_{\mathrm{halo}}$, this vertical offset directly measures the ratio of effective feedback energy coupled to the ISM and CGM per unit mass of black hole growth versus star formation ($\epsilon_{\mathrm{BH,eff}} / \epsilon_{\star,\mathrm{eff}} \sim 500$; see also \citealt{Donahue&Voit2022} and \citealt{Voit2024}). Because star-forming and quiescent central galaxies overlap substantially across a range of halo masses $M_{\mathrm{halo}} \sim 10^{12}\,M_\odot$ rather than occupying completely disjoint halo-mass regimes, this relative offset is less sensitive to mass-dependent halo-proxy systematics than is the power-law slope $b$. In modern cosmological simulations, stellar and AGN subgrid feedback efficiencies are often calibrated independently against separate observables (such as the SMHM and $M_{\mathrm{BH}}$--$\sigma$ relations; \citealt{Vogelsberger2014,Schaye2015,Weinberger2018}); requiring models to simultaneously reproduce an unbroken scaling and this $\sim 500\times$ effective-feedback ratio on a common halo-mass scale provides a stringent joint test of how both feedback channels couple to halo gas.

\subsection{Centrals and Satellites}
\label{central discussion}

As noted in Section~\ref{central method}, our centrality flag identifies confirmed centrals rather than positively separating satellites from unflagged galaxies, and fitting the full sample yields nearly identical parameters to the central-only fit (Section \ref{results}). This agreement highlights a useful feature of using $M_{\mathrm{GCS}}$ as a halo-mass proxy. For a satellite galaxy, $M_{\mathrm{GCS}}$ does not measure the virial mass of the parent group or cluster halo; instead, because many globular clusters form relatively early and sit deeper in the potential well where they are less readily stripped than the outer dark matter halo, $M_{\mathrm{GCS}}$ in a satellite should track something closer to a peak or pre-infall subhalo mass \citep[e.g.,][]{Hudson2014,Wetzel2013,Montero-Dorta2024}. For testing self-regulation, a subhalo or pre-infall mass scale is more physically relevant than the current host halo mass, because a galaxy's internal feedback history is governed primarily by its own potential well prior to strong environmental processing.

Two features of the non-central population in Figure~\ref{combined} are worth some reflection. First, massive quiescent satellites fall squarely on the relation traced out by their central brethren; these likely reflect regulation established before or during early infall, when their own halos and central black holes were actively governing gas cooling. Second, the low-mass quiescent galaxies visible in Figure~\ref{combined} are naturally interpreted as systems originally regulated by stellar feedback whose gas reservoirs were subsequently depleted or stripped in denser environments, leaving them quenched at lower masses without needing to invoke AGN feedback as the primary driver \citep[e.g.,][]{Bell2008,Wetzel2013}.

More broadly, these results align with observational and theoretical work showing that satellite quenching and gas removal take time to operate on infalling satellites. Many satellites retain CGM or ISM gas (visible as X-ray coronae around many satellites in clusters; \citealt{Sun2007}) and/or continue to form stars for extended timescales after entering a group or cluster \citep[e.g.,][]{Pasquali2010,Wetzel2013}. When environmental gas loss is gradual or incomplete, satellites can continue to reflect their prior internal regulation, only departing significantly from the central scaling relation when stripping or tidal mass loss becomes severe.

\section{Summary and Conclusions}
\label{summary}

The study of feedback in galaxies via observed scaling relations has historically been siloed into independent investigations of stellar and AGN feedback due to the limited valid or practical mass ranges of their $M_{\mathrm{halo}}$ proxies, with crosstalk limited to qualitative similarities in the SMHM and BHHM relations on one hand, and the quenching effects of AGN feedback on the other.  To bridge this gap, we use globular cluster system masses ($M_{\mathrm{GCS}}$) as a near-linear halo-mass proxy across roughly four orders of magnitude in $M_{\mathrm{halo}}$ to examine how galaxies, dark matter halos, and central supermassive black holes (SMBHs) connect across star-forming and quiescent regimes. Under a baryon-lifting view, if both star-forming and quiescent populations regulate halo gas against the same underlying potential well, differing mainly in their overall efficiency of converting mass into a baryon-lifting effect, their respective mass tracers should follow a common power-law scaling separated by a characteristic efficiency offset. Testing this directly, we find that a four-parameter unified power law with slope $b \approx 1.13$ and an offset of $d \approx 2.68\,\mathrm{dex}$ ($\sim 500\times$) between stellar and black hole feedback proxies achieves comparable statistical performance to separate regime-specific fits while remaining robust to including non-central galaxies, making its conceptual simplicity and unified framing an advantage even where statistical criteria are close.

This unified scaling is consistent with a baryon-lifting picture in which galaxies energetically self-regulate their circumgalactic media across mass scales, with the physical source of feedback mattering primarily through the net baryon-lifting work produced per unit mass of feedback tracer. In this view, the high-mass turnover or ``knee'' in the SMHM relation reflects a transition from stellar- to AGN-dominated regulation rather than a breakdown of energetic regulation itself, while non-central galaxies largely reflect pre-infall regulation and gradual rather than instantaneous environmental gas loss. Systematic differences between halo-mass proxies, modest departures from exact $M_{\mathrm{GCS}}$--$M_{\mathrm{halo}}$ linearity, parameter covariance, and mass-dependent cooling times all temper how tightly the power-law slope can be interpreted (on the order of $\pm 0.1$--$0.2$ in slope). Testing this framework further will benefit from targeted SMBH mass measurements in low-mass and star-forming systems, as well as checking whether cosmological simulations---whose stellar and AGN feedback prescriptions are often calibrated separately---reproduce both the unbroken scaling and the $\sim 2.68\,\mathrm{dex}$ offset simultaneously.

\begin{acknowledgments}

The authors graciously thank G. Mark Voit and Bryan A. Terrazas for their feedback, encouragement, and wisdom in the early stages of this work. This project would not have been conceivable without the groundwork they have set in the literature. We also thank all those whose questions and comments at various stages pushed this work forward technically and conceptually, especially Cayenne Matt, Emily Rauscher, Oleg Y. Gnedin, Adam Smercina, Katherine E. Whitaker, Min Yun, Yuan Li, Houjun Mo, and Ryan C. Hickox.

A. VanZandt expresses deep personal gratitude to Margot Lapeer and Benjamin N. Velguth, whose unfailing personal support and intellectual companionship were instrumental in the completion of this work.

This research has made use of the NASA/IPAC Extragalactic Database (NED), which is funded by the National Aeronautics and Space Administration and operated by the California Institute of Technology. This research has also made use of the Science Explorer (SciX) / NASA's Astrophysics Data System Bibliographic Services, funded by NASA under Cooperative Agreement 80NSSC25M7105. 

We have likewise made extensive use of the Simbad astronomical database \citep{simbad}

During the preparation of this work, the authors used a large language model (Google Gemini) to assist with reference finding, prose editing, text organization, and phrasing suggestions for portions of the manuscript. All scientific analysis, model fitting, interpretations, and literature checks were conceived, performed, and verified by the authors, who take full responsibility for the accuracy and integrity of the content.

This work made use of the following software packages: \texttt{astropy} \citep{astropy:2013,astropy:2018,astropy:2022,astropy_24340642}, \texttt{matplotlib} \citep{Hunter:2007}, \texttt{numpy} \citep{numpy}, \texttt{pandas} \citep{mckinney-proc-scipy-2010,pandas_23955360}, \texttt{python} \citep{python}, \texttt{scipy} \citep{2020SciPy-NMeth,scipy_22051554}, \texttt{astroquery} \citep{2019AJ....157...98G,astroquery_17163526}, \texttt{corner.py} \citep{corner-Foreman-Mackey-2016,corner.py_21208839}, \texttt{Cython} \citep{cython:2011}, \texttt{emcee} \citep{emcee-Foreman-Mackey-2013,emcee_10996751}, \texttt{h5py} \citep{collette_python_hdf5_2014,h5py_7560547}, \texttt{scikit-image} \citep{scikit-image}, \texttt{seaborn} \citep{Waskom2021}, \texttt{plotly} \citep{plotly} and \texttt{uncertainties} \citep{uncertainties}.

Software citation information aggregated using \texttt{\href{https://flatironinstitute.github.io/software-citation-station/}{The Software Citation Station}} \citep{software-citation-station-paper,software-citation-station-zenodo}.
\end{acknowledgments}

\pagebreak
\bibliographystyle{aasjournal}
\bibliography{refs}

\end{document}